\documentclass[11pt,a4paper]{article}
\usepackage[utf8]{inputenc}
\usepackage{amsmath}
\usepackage{mathtools}
\usepackage{a4wide}
\usepackage{cite}
\usepackage{authblk}

\title{Minkowski spacetime and non-Ricci-flat compactification\\ in heterotic supergravity}
\author{Takanao Tsuyuki\footnote{tsuyuki@cc.kogakuin.ac.jp}}
\affil{\small \textit{Academic Support Center, Kogakuin University, Hachioji, Tokyo 192-0015, Japan}}
\date{}

\begin{document}

\maketitle
\begin{abstract}
We compactify the ten-dimensional spacetime in heterotic supergravity leaving four-dimensional Minkowski spacetime. We search for nonsupersymmetric, non-Ricci-flat solutions of the equations of motion with the quadratic curvature term. By assuming that the extradimensional spaces are products of 2-manifolds, three types of solutions are found. They are $S^2\times T^2 \times H^2/\Gamma$, $S^2\times H^2/\Gamma \times H^2/\Gamma$, and $S^2\times S^2 \times H^2/\Gamma$, where $H^2/\Gamma$ denotes a compact hyperbolic manifold. The metrics can be written explicitly, and they can be applied to phenomenology.
\end{abstract}

\section{Introduction}
There are several problems in the Standard Model of particle physics: It does not contain gravity, and it cannot explain the origin of gauge groups or the number of generations. These problems would be explained by more fundamental theories, and prominent candidates for them are superstring theories. Superstring theories reside in ten-dimensional spacetime. Dimensions beyond four have not been observed \cite{Lee:2020zjt,Tan:2020vpf,Sirunyan:2020fwm,Zyla:2020zbs}, so the additional six dimensions need to be small enough. In the context of string phenomenology, the six-dimensional space is often assumed to be a Calabi-Yau manifold \cite{Candelas:1985en} or a toroidal orbifold \cite{Dixon:1985jw,Dixon:1986jc}. These spaces are Ricci flat, so it is easy to show that they satisfy the equations of motion.

The purpose of this paper is to find nonsupersymmetric, non-Ricci-flat compactifications in the supergravity theory as the low-energy limit of the heterotic superstring theory \cite{Gross:1984dd}.
Without supersymmetry, we need to solve equations of motion. We assume that the ten-dimensional spacetime is a direct product of Minkowski spacetime and three 2-manifolds. The advantage of 2-manifolds is that the condition for Green-Schwarz anomaly cancellation mechanism \cite{Green:1984sg} can be satisfied \cite{Witten:1984dg} without assuming equality between curvatures and gauge field strengths (``standard embedding''). In addition, such submanifolds allow us to suppose a Freund-Rubin-like configuration for gauge fields \cite{Freund:1980xh}. In the heterotic supergravity Lagrangian, the necessity for a quadratic curvature term was shown later in Ref. \cite{Bergshoeff:1989de}. Obtaining four-dimensional theories with non-Ricci-flat manifolds has, of course, a long history \cite{Freund:1980xh,Volkov:1980kq,Arefeva:1985yn,Kaloper:2000jb,Kehagias:2000dga,Townsend:2003fx,Ohta:2003pu,Chen:2003dca,Bytsenko:2005rm,Orlando:2010kx,Bars:1985ia,Govindarajan:1986kb,Castellani:1986rg,Israel:2004cd,Fernandez:2008wa,Lechtenfeld:2010dr,Brown:2013mwa,Lust:2020npd,Basile:2020mpt}. Our study differs from previous works for the quadratic curvature term and the above \textit{Ans\"{a}tze}.

To search for explicit solutions, we assume the two-dimensional submanifolds are spaces of constant curvature. The six-dimensional manifold solutions that we find here are $S^2\times T^2 \times H^2/\Gamma$, $S^2\times H^2/\Gamma \times H^2/\Gamma$ and $S^2\times S^2 \times H^2/\Gamma$, where $S^2$, $T^2$, and $H^2/\Gamma$ are a two-dimensional sphere, a torus, and a compact hyperbolic manifold \cite{Arefeva:1985yn,Orlando:2010kx}, respectively. By the flux quantization condition, a finite number of solutions for the curvature of the first $S^2$ are found.

This paper is organized as follows. In Sec. 2, we describe the Lagrangian and field equations that we solve in this paper. In Sec. 3, explicit configurations of gauge fields and curvature tensors that satisfy the field equations are shown. In Sec. 4, we summarize the results of this paper.

\section{Lagrangian and equations of motion}

We consider the bosonic part of the Lagrangian for heterotic supergravity \cite{Bergshoeff:1989de,Lechtenfeld:2010dr}
\begin{align}
L=\sqrt{-g}e^{-2\phi}\left[R+4(\nabla \phi)^2-\frac{1}{12}H_{MNP}H^{MNP}+\frac{\alpha'}{8}R_{MNPQ}R^{MNPQ}-\frac{\alpha'}{8}\textrm{tr}(F_{MN}F^{MN})\right],
\end{align}
where $g$ is the determinant of the metric $g_{MN}$ and the indices run $M,N,P,Q =0,\ldots,9$. $R_{MNPQ}$ is the Riemann tensor, and the Ricci scalar $R$ and the Ricci tensor $R_{MN}$ are defined as  $R= g^{NQ}R_{NQ} = g^{MP}g^{NQ} R_{MNPQ}$. The gauge field strength $F_{MN}$ is
\begin{align}
F_{MN} &= \partial_{M}A_{N}-\partial_N A_M+[A_M, A_N], \\
F_{MN} &= F_{MN}^A T^A, \ A_M = A_M^A T^A, \ \textrm{tr}(T^A T^B) =\delta^{AB}, 
\end{align}
where $A_M^A$ is a gauge field and $T^A$ is a generator of gauge groups $E_8\times E_8$ or $SO(32)$.

The equation of motion for the three-form field $H$ is
\begin{align}
\partial^M(e^{-2\phi}H_{MNP})=0. \label{ehmn}
\end{align}
We take dilaton $\phi$ constant and $H$ to vanish:
\begin{align}
\partial_M \phi=0,\quad H_{MNP}=0, \label{edp}
\end{align}
and then the equation of motion for $H$ (\ref{ehmn}) is satisfied. Such a configuration with unbroken supersymmetry leads to Ricci-flat compactification \cite{Candelas:1985en}. In this paper, we do not assume low-energy supersymmetry. The other equations of motion become
\begin{align}
R+\frac{\alpha'}{8}R_{MNPQ}R^{MNPQ}-\frac{\alpha'}{8}\textrm{tr}(F_{MN}F^{MN}) &=0,\label{eral}\\
R_{MN}+\frac{\alpha'}{4}R_{MPQR}{R_N}^{PQR} -\frac{\alpha'}{4}\textrm{tr}(F_{MP}{F_N}^{P}) &=0, \label{ermn}\\
\nabla_M F^{MN}+[A_{M},F^{MN}] &=0. \label{efm}
\end{align}
Equation (\ref{eral}) can be made much simpler. By multiplying Eq. (\ref{ermn}) with $g^{MN}$, we obtain
\begin{align}
R+\frac{\alpha'}{4}R_{MNPQ}R^{MNPQ}-\frac{\alpha'}{4}\textrm{tr}(F_{MN}F^{MN}) &=0.
\end{align}
Comparing with Eq. (\ref{eral}), we find
\begin{align}
R=0. \label{er0}
\end{align}
This is a result of the constant dilaton and vanishing $H$ (\ref{edp}). We use this equation instead of the dilaton equation (\ref{eral}). Vanishing of the Ricci scalar does not mean that the ten-dimensional manifold is Ricci-flat $R_{MN}=0$, as we will see in the next section.

In addition to the above equations of motion, the curvature form and the gauge field have to satisfy
\begin{align}
0=dH=\frac{\alpha'}{4}(\textrm{tr} R\wedge R - \textrm{tr} F\wedge F) \label{edh}
\end{align}
for the Green-Schwarz anomaly cancellation mechanism \cite{Green:1984sg}. To satisfy this condition, standard embedding $R_{MNab}=F^A_{MN}T^A_{ab}$  has often been assumed. Such a relation, however, cancels the second and the third terms of Eq. (\ref{ermn}), and only Ricci-flat solutions $R_{MN}=0$ are allowed. We need
\begin{align}
{R_{MNab}} \neq F^A_{MN}T^A_{ab} \label{erf}
\end{align}
for some components to find nontrivial solutions.

\section{Compactification}
\subsection{Ans\"{a}tze}
We are going to solve Eqs. (\ref{ermn}), (\ref{efm}), (\ref{er0}), and (\ref{edh}). We assume that the ten-dimensional manifold is a product of four manifolds:
\begin{align}
M^{10}=M_0\times M_1\times M_2\times M_3, \label{em10}
\end{align}
where $M_0$ is the four-dimensional Minkowski spacetime and $M_i\ (i=1,2,3)$ are two-dimensional spaces of constant curvature. The metric of $M^{10}$ is block diagonal and depends only on the coordinates of corresponding manifolds $g_{mn}^{(i)}=g_{mn}^{(i)}(x^{(i)})$, where the indices $m$ and $n$ are tangent to $M_i$. The nonzero Riemann tensor components are then
\begin{align}
R_{mnpq}^{(i)} &=\lambda_i (g_{mp}^{(i)}g_{nq}^{(i)}-g_{mq}^{(i)}g_{np}^{(i)}), \label{erm}
\end{align}
where $\lambda_i$ is a constant sectional curvature. The other components with indices of $M_0$ (like $R_{0123}$)  or mixed manifolds (like $R_{4568}$) are zero.

For the gauge field strength $F_{MN}^A$, we assume that it is also block diagonal for $M$ and N, and nonzero only for $U(1)$ components $(A=1,2,\ldots,12)$. The range of $A$ is determined to leave the gauge group $SU(5)$ in $E^8\times E^8\supset SU(5)\times U(1)^{12}$. 
For the nonzero components, we take Freund-Rubin-like configuration \cite{Freund:1980xh}
\begin{align}
F_{mn}^{A(i)} &=\sqrt{g_i}f^A_i \epsilon_{mn}^{(i)}, \label{efmn}
\end{align}
where $g_i=\textrm{det}(g_{mn}^{(i)})$, $f^A_i$ is a constant (sometimes called a flux density \cite{Brown:2013mwa}), and $\epsilon_{mn}^{(i)}$ is a Levi-Civita symbol for $M_i$. Other components are set to zero. This configuration can satisfy the nonstandard embedding condition (\ref{erf}).

\subsection{Field equations and flux quantization}
By the above setups, we can show that the equation of motion for $F_{MN}$ (\ref{efm}) is  satisfied. Consider that $N=n$ is in the direction of the manifold $M_i$. The second term in Eq. (\ref{efm}) vanishes
\begin{align}
[A_M, F^{Mn}]=A_m^{A(i)} F^{Bmn(i)}[T^A, T^B]=0,
\end{align}
since $A^A$ is nonzero only for $U(1)$ components. The covariant derivative term becomes also zero for the Freund-Rubin configuration:
\begin{align}
\nabla_M F^{Mn} =\frac{1}{\sqrt{g_i}}\partial_m \left(\sqrt{g_i}\frac{f_i^A}{\sqrt{g_i}}\epsilon^{mn}\right)T^A=0.
\end{align}

Our \textit{Ans\"atze} Eqs. (\ref{erm}) and (\ref{efmn}) can also satisfy the condition for the Green-Schwarz mechanism, Eq. (\ref{edh}). Components of the first term are zero \cite{Witten:1984dg}:
\begin{align}
{R_{[MN}}^{RS}R_{PQ]RS} &= \sum_{i=1}^3 {R_{[MN}^{(i)}}^{rs}R_{PQ]rs}^{(i)} =0.
\end{align}
This is because $\{M,N,P,Q\}$ cannot be all different in each two-dimensional manifold $M_i$. For the same reason, the second term in Eq. (\ref{efmn}) is also zero if $M,N,P$ and $Q$ are in the same $M_i$. For different indexes $m$ and $n$ tangent to $M_i$ and $p$ and $q$ tangent to $M_j$ ($i\neq j$), we need 
\begin{align}
F_{[mn}^AF_{pq]}^A &= \frac{1}{3}\sum_A \sqrt{g_ig_j} f^A_if^A_j = 0.
\end{align}
It can be satisfied if $f_i^A$ do not overlap for $A$:
\begin{align}
f^A_if^A_j = 0 \ (i\neq j,\ \textrm{no sum over } A ). \label{efafa}
\end{align}

The equations of motion to be solved are Eqs. (\ref{ermn}) and (\ref{er0}). They are reduced to
\begin{align}
\lambda_i +\frac{\alpha'}{2}\left(\lambda_i^2 -\sum_A(f_i^{A})^2\right) &= 0, \label{elamb}\\
\sum_{i=1}^3 \lambda_i &= 0. \label{esum}
\end{align}
From Eq. (\ref{elamb}), we see that $\lambda_i$ can be positive or negative. This is crucial to obtain non-Ricci-flat solutions. If the quadratic curvature term or gauge field were not, all $\lambda_i$ are nonpositive or non-negative, and Eq. (\ref{esum}) forces them to be zero.

The equations of motion (\ref{elamb}) and (\ref{esum}) have a trivial flat solution $M_1=M_2=M_3=T^2$ ($\lambda_1=\lambda_2=\lambda_3=0$ with no gauge fields, $f_1^A=f_2^A=f_3^A=0$). We are interested in nonflat solutions here. The manifold with $\lambda_i>0$ is a two-sphere $S^2$ and $\lambda_i<0$ is a compact hyperbolic manifold $H^2/\Gamma$ \cite{Arefeva:1985yn,Orlando:2010kx}.
For $M_i=S^2,H^2/\Gamma$, gauge field strength satisfies the flux quantization condition \cite{Witten:1984dg,Lust:2020npd}
\begin{align} 
\int_{M_i} F^A = \textrm{vol}(M_i) f_i^A = 2\pi n_{i}^A, \label{efv}
\end{align}
where vol$(M_i)$ is a volume of $M_i$ and $n_{i}^A$ is an integer. On the other hand, the Ricci scalar is also related to an integer by the Gauss-Bonnet theorem:
\begin{align} 
\int_{M_i}R=\textrm{vol}(M_i)2\lambda_i=4\pi\chi_i, \label{erv}
\end{align}
where $\chi_i$ is the Euler characteristic ($\chi_i=2$ for $S^2$ and a negative even number for $H^2/\Gamma$). Dividing Eq. (\ref{efv}) by Eq. (\ref{erv}), we find
\begin{align}
f_i^A=\frac{n_{i}^A}{\chi_i}\lambda_i. \label{efia}
\end{align}
Substituting it to the equation of motion (\ref{elamb}), we obtain
\begin{align} 
\lambda_i &= 
\begin{dcases}
\frac{1}{c_i-1}\frac{2}{\alpha'}  & (c_i\neq 1)   \\
0 & (c_i=1)
\end{dcases},  \label{elic}\\
c_i &\equiv \frac{1}{\chi_i^2}\sum_A (n_{i}^A)^2. \label{ecic}
\end{align}

\subsection{Solutions}

For simplicity, we put $\alpha'=2$ below. To find solutions for Eqs. (\ref{esum}), (\ref{elic}), and (\ref{ecic}), it is useful to derive the range of $\lambda_i$. In the region $c_i<1$ and $c_i>1$, $\lambda_i$ is monotonically decreasing with $c_i$. By the definition, $c_i\geq 0$ and the minimal $c_i>1$ is $\frac{5}{4}$; then
\begin{align}
\lambda_i \leq
\begin{dcases}
4 & (\textrm{for } S^2) \\
-1 & (\textrm{for } H^2/\Gamma)
\end{dcases}. \label{ila}
\end{align}
Without loss of generality, we can set $\lambda_1\geq \lambda_2\geq \lambda_3$. For nonflat solutions, Eq. (\ref{esum}) implies $\lambda_1>0$ and $\lambda_3<0$. We discuss three cases with zero, negative, and positive $\lambda_2$. We can find many discretized solutions, and two sample solutions for each case are summarized in Table~\ref{tl1}. We pick up some solutions below.

\begin{table}[b]
\centering
\caption{Sample solutions for the equations of motion (\ref{elamb}) and (\ref{esum}), satisfying the conditions for anomaly cancellation (\ref{efafa}) and flux quantization (\ref{efia}). The sectional curvature $\lambda_i$ and the flux density $f_i^A$ are defined in Eqs. (\ref{erm}) and (\ref{efmn}), and $\chi_i$ is the Euler characteristic ($\chi_1=2$).
$\lambda_i$ and $f_i^A$ are proportional to $2/\alpha'$, and $m$ and $n$ are arbitrary natural numbers.}
\label{tl1}
\begin{tabular}{rrrrrl} \hline \hline
$\lambda_1$   & $\lambda_2$   & $\lambda_3$ & $\chi_2$ & $\chi_3$ & Nonzero $f_i^A$                                        \\ \hline
$1$           & $0$           & $-1$ & $0$   & $-2n$ & $f_1^1=f_1^2=1 $                                    \\
$4$           & $0$           & $-4$ & $0$   & $-2n$ & $f_1^1=4,f_1^2=f_3^3=f_3^4=f_3^5=2$                 \\
$2$           & $-1$          & $-1$ & $-2m$ & $-2n$ & $f_1^1=2,f_1^2=f_1^3=1$                             \\
$4$           & $-2$          & $-2$ & $-2m$ & $-2n$ & $f_1^1=4, f_1^2=2, f_2^3=f_2^4=f_3^5=f_3^6=1$       \\
$\frac{2}{3}$ & $\frac{1}{3}$ & $-1$ & $2$   & $-2n$ & $f_1^1=1,f_1^2=\frac{1}{3}, f_2^3=\frac{2}{3}$      \\
$4$           & $4$           & $-8$ & $2$   & $-4n$ & $f_1^1=f_2^3=f_3^5=4, f_1^2=f_2^4=f_3^6=2, f_3^7=6$
\\ \hline \hline
\end{tabular} 
\end{table}

\subsubsection{Case (i): $M_1=S^2, M_2=T^2$ and $M_3=H^2/\Gamma$}
This manifold corresponds to $\lambda_2=f_2^A=0$. The condition of vanishing Ricci scalar (\ref{esum}) is \begin{align}
\frac{1}{c_1-1}+\frac{1}{c_3-1} = 0.
\end{align}
It has four solutions,
\begin{align}
(c_1, c_3) = \left(\frac{5}{4}, \frac{3}{4}\right),\ \left(\frac{3}{2}, \frac{1}{2}\right), \left(\frac{7}{4}, \frac{1}{4}\right),\ (2, 0),\end{align}
corresponding to
\begin{align}
\lambda_1&=-\lambda_3=4,2,\frac{4}{3},1.
\end{align}
One of the solutions $(\lambda_1, \lambda_3)=(1,-1)$ can be realized if the nonzero parameters are (we choose $f_i^A \geq 0$)
\begin{align}
n_1^1=n_1^2=2,\ f_1^1=f_1^2=1.
\end{align}
The number of nonzero $f_i^A$ is two, and it is minimal for case (i). 
The condition for anomaly cancellation Eq. (\ref{efafa}) is satisfied.

\subsubsection{Case (ii): $M_1=S^2$ and $M_2=M_3=H^2/\Gamma$}
This manifold corresponds to $\lambda_2<0$. The inequality (\ref{ila}) and Eq. (\ref{esum}) give
\begin{align}
4\geq \lambda_1 = -\lambda_2-\lambda_3 \geq 2.
\end{align}
There are only two solutions for $\lambda_1$:
\begin{align}
\lambda_1=4, 2.
\end{align}
For $\lambda_1=2$, the others are $\lambda_2=\lambda_3=-1$. This case requires three $f_i^A$ to be nonzero, and we have checked that this is minimal. 
For $\lambda_1=4$, other parameters $\lambda_i$ and $f_i^A$ can be calculated as in other cases.

\subsubsection{Case (iii): $M_1=M_2=S^2$ and $M_3=H^2/\Gamma$}
The ranges of the sectional curvatures are
\begin{align}
4&\geq\lambda_1\geq (\lambda_1+\lambda_2)/2 =- \lambda_3/2 \geq 1/2
\end{align}
Then eight values of $\lambda_1$ are possible:
\begin{align}
\lambda_1=4, 2, \frac{4}{3}, 1,\frac{4}{5}, \frac{2}{3}, \frac{4}{7}, \frac{1}{2}.
\end{align}
The minimal number of nonzero $f_i^A$ is three for the case $(\lambda_1, \lambda_2, \lambda_3)=(\frac{2}{3},\frac{1}{3},-1)$. Another nontrivial solution $(\lambda_1,\lambda_2,\lambda_3)=(4,4,-8)$ can be realized with
\begin{align}
c_1=c_2=\frac{5}{4}=\frac{2^2+1^2}{2^2},\ c_3=\frac{7}{8}=\frac{(3n)^2+(2n)^2+n^2}{(-4n)^2}, \label{ec1c2}
\end{align}
where $n$ is an arbitrary natural number. The Euler characteristic $\chi_3=-4n$ is limited to a multiple of four. Nonzero flux densities $f_i^A$ are calculated by Eqs. (\ref{efia}) and (\ref{ec1c2}) as 
\begin{align}
f_1^1 = f_2^3=f_3^5=4,\ f_1^2=f_2^4=f_3^5=2,\ f_3^7=6.
\end{align}
The anomaly cancellation condition Eq. (\ref{efafa}) is satisfied.

\section{Conclusions}
We found a new set of nontrivial solutions for the equations of motion in heterotic supergravity. We have assumed the extra six-dimensional space is a product of 2-manifolds (\ref{em10}). Such a space has two advantages: The Green-Schwarz mechanism can be realized without assuming the standard embedding, and the gauge field can take Freund-Rubin-like configuration (\ref{efmn}). By the quadratic curvature term in the Lagrangian and the nonstandard embedding, the sectional curvature $\lambda_i$ can be positive or negative [see Eq. (\ref{elamb})], leading to non-Ricci-flat solutions. The solutions are summarized in Table~\ref{tl1}. 

In these solutions, the curvatures are fixed by the equations of motion and discretized by the flux quantization condition. They cannot be continuously changed, so the solutions are expected to be stable. In contrast with Ricci-flat cases, there is no freedom to multiply curvatures since the quadratic curvature term is present, and such stability was pointed out in Ref. \cite{Lechtenfeld:2010dr} (the stability of compactification on the products of Einstein manifolds was also discussed in Ref. \cite{Brown:2013mwa} with a different action). It would be interesting to examine whether the solutions remain stable with the inclusion of higher-curvature terms.

Our compactifications realize four-dimensional Minkowski spacetime, so they can be applied for phenomenology. For example, if nonzero $U(1)$ fluxes are suitably embedded in $E^8\times E^8$, the gauge symmetry $SO(10)$ can be left, which is ideal for grand unification with massive neutrinos \cite{Georgi:1974my,Fritzsch:1974nn}. We know explicit metrics and gauge field components in our solutions. They will enable us to calculate not only topological numbers such as the number of generations, but also other continuous parameters like Yukawa couplings in four-dimensional theories. We would further study these phenomenological aspects in the future.

\section*{Acknowledgments}
The author thanks Shun’ya Mizoguchi for useful discussions and comments.

\end{document}